# Cross-Media Scientific Research Achievements Retrieval Based on Deep Language Model

Benzhi Wang, Meiyu Liang*, Feifei Kou and Mingying Xu

（School of Computer Science, Beijing Key Laboratory of Intelligent Telecommunication Software and Multimedia, Beijing University of Posts and Telecommunications, Beijing 100876, China）

Abstract  Science and technology big data contain a lot of cross-media information.There are images and texts in the scientific paper.The s ingle modal search method cannot well meet the needs of scientific researchers.This paper proposes a cross-media scientific research achievements retrieval method based on deep language model (CARDL).It achieves a unified cross-media semantic representation by learning the semantic association between different modal data, and is applied to the generation of text semantic vector of scientific research achievements, and then cross-media retrieval is realized through semantic similarity matching between different modal data.Experimental results show that the proposed CARDL method achieves better cross-modal retrieval performance than existing methods.
Key words science and technology big data ; cross-media retrieval; cross-media semantic association learning; deep language model; semantic similarity

With the rapid development of information technology, the scale of science and technology big data is growing rapidly [1]. The traditional data query and retrieval method [2] is no longer suitable for the information retrieval needs of scientific researchers. How to find and obtain valuable information from the massive cross-media technology big data is a hot research topic. In the cross-media technology big data, there are scholars, institutions, papers, projects, patents and other data, which are in a huge number and various kinds. Science and technology big data brings people a lot of information and challenges, and researchers need to find valuable information from the massive cross-media big data. The existing big data retrieval technology cannot be directly applicable to science and technology big data, and suitable methods for science and technology big data retrieval query task are studied. Science and technology big data has many media types of scientific and technological resources. How to transform these cross-media scientific and technological resources into a unified semantic representation is an urgent problem to be solved. This paper proposes a cross-media scientific research achievement retrieval method (CARDL) based on the deep language model, to realize the efficient and accurate cross-media retrieval of massive scientific and technological achievement data.The main contributions of this article are as follows:

(1) For the deep language model of scientific research achievements, the generation of the text semantic vector of scientific research results is realized, and its depth semantics is used to enhance the characterization ability of text data.

(2) Cross-media scientific research achievement retrieval method based on deep language model is proposed. A unified intermedia semantic representation is obtained through the learning of semantic correlation between different modal data, and cross-media retrieval is realized through the semantic similarity matching between different modal data.

## 1 Related work

Scientific research achievements include scientific and technological resources from different fields and in different modes. Literature [3][4][5][6] propose methods to transform different cross-media data into unified features, and the main method is to map data from different modalities to a unified feature subspace through cross-collaborative learning [7] according to different modal features. Traditional feature extraction methods include one-hot, TF-IDF, etc. One-hot code [8][9] N states are encoded with N-bit state registers, each with independent register bits. TF-IDF [10] Assess the importance of a single word to a file set or a file in a corpus. The importance of a word increases proportional with the number of times it appears in the file, but also decreases inversely with the frequency that it appears in the corpus. Although both methods also have novel applications [11], They do not take into account the occurrence order of symbols in the text set, do not reflect the location information of symbols, and rely heavily on corpora, etc.

Google earlier opened-source word2vec, a tool for word vector computing, Word2vec [12][13][14]. It can be efficiently trained on orders of millions of dictionaries and hundreds of millions of datasets, and the word vector obtained by the tool can measure the similarity between words and words well. The Word2vec has two basic models, namely, CBOW and skip-gram. CBOW（continuous Bag-of-Words Model）[15][16] Prediction of the current word by context, skip-gram [17] Based on the current word prediction context, the disadvantage of the two ways mentioned above do not contain location information.

The BERT, as proposed by Google [15] Essentially learning a good feature representation for words by running self-supervised learning methods based on massive corpora. Self-supervised learning is supervised learning running on data without manual annotation [19]. In a specific NLP task [20][21], features can directly represent BERT as a word embedding feature for the task.BERT provides a model for transfer learning from other tasks [22][23], The model can be used as a feature extractor after fine-tuning or being fixed by the task. Currently, sources and models for BERT are open source on online platforms [24]。

Document [25][26][27][28] A cross-modal query method is proposed for feature similarity matching [29] by mapping the technological resources with different modes to a unified feature space. Semantic fusion of the different modal data [30][31] The methods to obtain unified semantic vectors include typical correlation analysis (CCA) [32], isomorphic spatial feature learning method (CSL), etc.With the popularity of deep learning technologies [33][34], Deep neural networks are widely used in the field of cross-modal search. document [35] A deep tensor typical correlation analysis (DTCCA) is presented. This is a method for learning complex nonlinear transformations [36] of multimodal data, such that the obtained representation is higher-order linearly dependent.

Cross-modal hashing (CMH) is widely used in similarity search due to its advantages of low storage cost and fast query speed.However, almost all existing CMH methods are based on hand-crafted features that may not be fully compatible with the hash code learning process.Therefore, existing CMH methods with manual features may not achieve satisfactory performance.document [37] A new cross-modal hash method is proposed, which integrates feature learning and hash code learning into the same framework. These methods are able to establish unified semantics between paired images and texts, but ignore possible correlations in the same space.

## 2. Deep semantic feature learning of scientific research achievement data

This paper optimizes the BERT deep language model based on the scientific and technological achievements data, and performs the deep semantic feature learning of the scientific and technological achievements data based on the optimized deep language model.Converts text data vectors into a specific length.Based on this, Google's BERT pre-training model was fine-tuned using the research corpus to make it more suitable for research data.The last hidden layer output vector of the convolutional neural network in a binclassification task of calculating image similarity was used as the initial feature of the image.

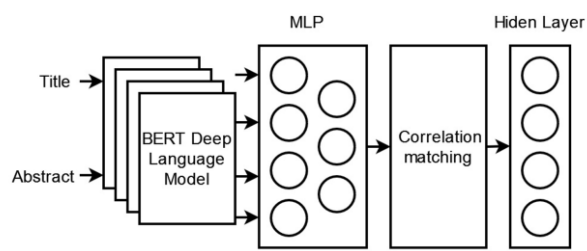

Figure 1 Framework diagram of deep semantic feature learning of scientific research achievement data

The BERT pre-trained model was fine-tuned using crawled technological achievement data.Fine-tuning of the selected task model is shown in Figure 1.The input part is the title of the paper and the sentence in the summary of the scientific research project. The title has a one-to-one correspondence with the abstract, and this correlation degree is used to fine-tune the deep language model.The titles and sentences go through the BERT model, by pooling the output vectors X with Y.Joint training with X, Y, | X-Y |, max (X, Y) to adjust the last

two hidden layers of BERT model to fit the corpus of scientific research results.

## 3. Cross-Media Scientific Research Achievement Search Method Based on Deep Language Model (CARDL)

The data of cross-media scientific research achievements mainly includes two modalities, image and text. The research objects of this paper mainly focus on the retrieval between the text and image data of scientific and technological achievements.The title and keywords of the technology paper itself correspond to the image.For the same keywords, some relevant sets of images can be constructed.Negative samples were taken using images from papers in different fields.In view of the characteristic ogeneity and semantic gap between different modal data of scientific and technological achievements, this paper proposes a cross-media scientific research retrieval method based on deep language model. The algorithm framework is shown in FIG. 2. By learning between different modal data of scientific and technological achievements, mapping different modal data to a unified semantic vector space, and then realizing cross-media search of scientific and technological achievements through semantic similarity matching.

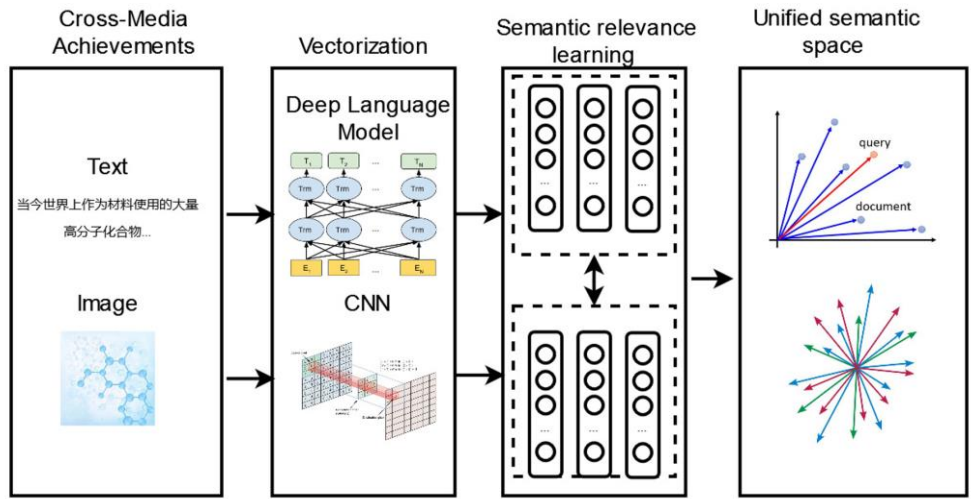

Figure 2 Framework diagram of the cross-media scientific research achievement retrieval method based on the deep language model

For text modality, a fixed-dimensional text feature vector is generated by the fine-tuned Chinese pre-training model BERT, followed by a multi-layer perceptual network (MLP).For image modalities, it is first processed into 256 * 256 to generate the image semantic convolution feature vector via a deep convolutional neural network (CNN) and an MLP.After the learned text and image semantic features were learned across media semantic associations, the similarity scores for both vectors were calculated using cosine similarity.

The objective loss function for cross-media semantic association learning of scientific and technological achievements data is shown in Equation (1), and mainly includes two parts, the loss function corresponding to labels for all images in the new space and the loss function corresponding to all text in the new space.

$$\mathcal{L}_{\text{total}} = \mathcal{L}_{I2T} + \mathcal{L}_{T2I} \quad （1）$$

$\mathcal{L}_{I2T}\mathcal{L}_{T2I}\mathcal{L}_{I2T}\mathcal{L}_{T2I}$Among them, the loss of picture and corresponding keyword and title in batch, the loss of keyword and corresponding picture, and both are cross entropy, and the optimization target is the image corresponding to the original space, and the text still have a high similarity in the new space.

$$\mathcal{L}_{I2T,T2I} = -\frac{1}{n}\sum_{i=1}^{n}\frac{1}{m}\sum_{j=1}^{m}\left(y_{ij} \cdot \log p_{ij}\right) \quad （2）$$

The distribution value of the i th data in the image and the j th tag is, calculating the comprehensive loss of the corresponding m labels in n samples.$p_{ij}$

After obtaining a cross-media unified semantic representation through cross-media semantic association learning, the cosine similarity is used to perform cross-media semantic similarity matching between image and text modes, further improving the search ability in large-scale data while ensuring high accuracy.Cosine distance calculation is well supported on existing search engines, it is calculated faster, and the semantic similarity calculation is calculated as follows:

$$\text{sim}(x, y) = \frac{x \cdot y}{\|x\| \cdot \|y\|} = \frac{\sum_{i=1}^{d_{x,y}} x_i \times y_i}{\sqrt{\sum_{i=1}^{d_{x,y}} (x_i)^2} \times \sqrt{\sum_{i=1}^{d_{x,y}} (y_i)^2}} \quad (3)$$

$xy$ It represents text features in unified semantic space and image features in unified semantic space. The overall process of the cross-media deep language model is shown in Table 1.

Table 1 Intermedia research method based on deep language model

| |
| --- |
| Algorithm: Intermedia scientific research achievement retrieval method based on deep language model |
| Input: Text and image set Q, the number of search results k |
| Output: Cross-media search results |
| (1) For each image in the text image collection: |
|     1) Image features were computed using the CNN |
|     2) Map image features to unified spatial semantic vectors |
| (2) Do the following for each text in the text image collection: |
|     1) Text features were computed using a deep language model |
|     2) Map text features to unified spatial semantic vectors |
| (3) Semantic vector similarity was calculated and matched between the features of all image-text pairs in the text and image collection in a unified semantic space |
| (4) Sort according by semantic similarity |
| (5) Returns the Top-k cross-media search results |

Table 2 Dataset

| Data types | source | total |
| --- | --- | --- |
| project | The National Natural Science Fund | 107847 |
| achievements in scientific research | The National Natural Science Fund | 1462751 |
| Scientific research scholar | The National Natural Science Fund | 428365 |
| thesis | Know the net | 782642 |
| The paper corresponds to the image | Know the net | 103589 |

## 4 Experiments and analysis

### 4.1 Dataset and evaluation indicators

To verify the query method of cross-media scientific research results proposed in this paper, we climbed the scientific research results data, and the data statistics are shown in Table 2. This paper trains the deep speech language model using the name of the project and the correspondence between abstracts.

Table 3 Paper images corresponding to the datasets

| Data types | Total number of papers | Total image |
| --- | --- | --- |
| organic chemical industry | 2173 | 4916 |
| biomedicine | 1754 | 4481 |
| Architectural Science and Engineering | 461 | 1364 |

The CNKI paper and corresponding images were used to verify the effectiveness of this method, and the paper images from organic chemistry, biomedicine, architectural science and engineering were selected as the data. The performance evaluation index used is MAP (Mean Average Precision, average accuracy), which utilized in many cross-media retrieval works [38][39]. The MAP index values were calculated from the mean of the average accuracy (AP) of all queries. The larger the MAP value, the better the cross-media search performance of scientific and technological achievements. The calculation formula is as follows:

$$MAP = \frac{\sum_{q=1}^{Q} AP(q)}{Q} \quad (4)$$

$$AP@R = \frac{\sum_{r=1}^{R} P(r)\delta(r)}{R^{'}} \quad (5)$$

Where Q represents the number of queries, R is the number of search results returned, and P(r) represents the accuracy of the first r search results. If the r th search results and query are relevant, otherwise. $\delta(r)=1$ $\delta(r)=0$ $R^{'}$ Denote the number of search results associated with the query.

### 4.2 Experimental results and analysis

This paper verifies the performance of deep language model in the entity name extraction of scientific research data before and after the optimization. The accuracy of 77% and recall of 78% reaches the accuracy of 82%, and the accuracy of deep language model based on scientific and technological achievement data is improved by nearly 5%. Images were feature encoded by CNN, with 256 * 256 input, output as 64 D vector through CNN and MLP network, and the text was output as 64 D vector through BERT.

Table 4 Comparison of cross-media research retrieval performance

| algorithm | MAP@1 txt2img | MAP@5 txt2img | MAP@10 txt2img |
|---|---|---|---|
| CCA | 0.2593 | 0.3817 | 0.4769 |
| CMDN | 0.2759 | 0.3987 | 0.4880 |
| **OURS** | 0.3255 | 0.4491 | 0.5158 |

Table 5 Comparison of cross-media research retrieval performance

| algorithm | MAP@1 img2txt | MAP@5 img2txt | MAP@10 img2txt |
|---|---|---|---|
| CCA | 0.1972 | 0.3576 | 0.3988 |
| CMDN | 0.2395 | 0.3409 | 0.3829 |
| **OURS** | 0.2801 | 0.3953 | 0.4624 |

The performance comparison of different algorithms in the text search map and graph search text of cross-media scientific and technological achievements is shown in Table 4 and Table 5, respectively, and the average performance is shown in Figure 3. It can be seen that the traditional method improves both CCA and CMDN to a certain extent.

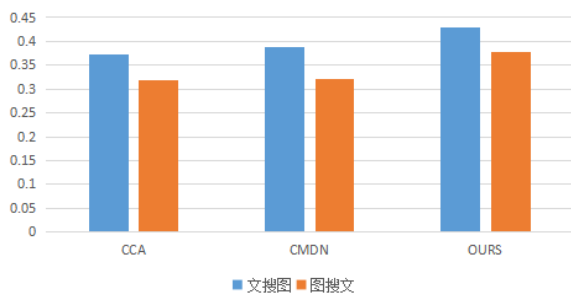

Figure 3 Average performance comparison of scientific research achievement retrieval methods across media

Because the CCA method focuses on the association relations of picture text pairs, it uses the association loss function as the optimization target, and ignores the semantic association relations of text images to internal elements. It cannot establish sufficient association relations for data with different same semantic modes, so the effect of both image search and text search is lower than that of CARDL method. Although CMDN also uses deep learning to train the unified semantic space, CARDL uses a custom task to pre-train for scientific research achievements to generate text features and image features, and the average MAP is 5% higher than CMDN in the retrieval results. After combining the BERT deep language model, CARDL eliminates certain differences in mode through the combination of deep semantic vector and image, and basically meets the retrieval needs of text search and text search in scientific research results.

## 5 Conclusion

Aiming at the deep language model of the scientific research results, this paper realizes the generation of the text semantic vector, and its deep semantics is used to enhance the characterization ability of the text data. Cross-media research results retrieval method based on deep language model, obtaining unified intermedia semantic representation by semantic association learning between different modal data, and achieving cross-media retrieval by semantic similarity matching between different modal data. Experimental results on a dataset of scientific and technological achievements verify the effectiveness of the proposed learning and retrieval method of intermedia scientific and technological achievements semantic representation.

arXiv preprint arXiv:2203.08615, 2022.

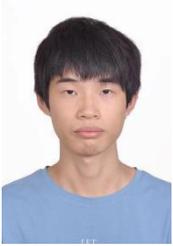

**Benzhi Wang**, born in 1997Master His main research interests include Data mining, information retrieval, machine learning.

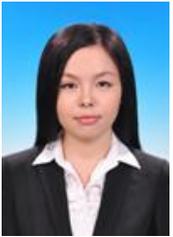

**Meiyu Liang**, was born in 1985. She is now an associate professor and master tutor at the School of Computer Science, Beijing University of Posts and Telecommunications.Her main research directions are artificial intelligence, data mining, multimedia information processing, and computer vision.

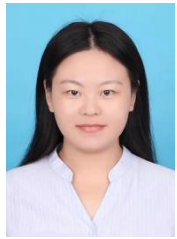

**Feifei Kou**, was born in 1989. She received her Ph.D.degree in School of Computer Science from Beijing University of Posts and Telecommunications, Beijing, China, in 2019. She ever did postdoctoral research in School of Computer Science from Beijing University of Posts and Telecommunications from 2019 to 2021. She is currently an lecturer in School of Computer Science (National Pilot software Engineering School), Beijing University of Posts and Telecommunications, Beijing, China.Her research interests include semantic learning, and multimedia information processing.

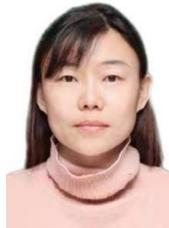

**Mingying Xu**, was born in 1987. Ph.D., CCF Member.Her main research interests include information retrieval, science and technology big data analysis, data mining.